# New Musical Interfaces and New Music-making Paradigms

**Sergi Jordà**

Music Technology Group, Audiovisual Institute, Pompeu Fabra University

Passeig de la Circumval·lació 8, 08003 Barcelona , Spain

sergi.jorda@iua.upf.es

**ABSTRACT**

The conception and design of new musical interfaces is a multidisciplinary area that tightly relates technology and artistic creation. In this paper, the author first exposes some of the questions he has posed himself during more than a decade experience as a performer, composer, interface and software designer, and educator. Finally, he illustrates these topics with some examples of his work.

**INTRODUCTION**

The conception and design of new musical interfaces is a burgeoning multidisciplinary area that relates very tightly, technological knowledge (sensors technology, computer programming, etc.) with a deep understanding of musicians culture. If new musical interfaces can be partially responsible for shaping some of the future music, these new musical paradigms should not be left to improvisation. We are not asserting that the design of new interfaces must follow any given tradition, but a wide knowledge of these, together with the personal beliefs and intuitions of each designer, should orient about what could be changed and what could be kept. New instruments designers have, in overview, two *traditions* in front of them: a millennial one, as old and rich as the history of music making, and a half-century old one, that of computer music. The first part of this paper briefly presents three topics (many others could have been chosen), discusses their current situations and suggests how fresh approaches can influence in the conception of new musical interfaces.

**Controlling at Macro and/or Micro Levels**

For centuries, Western music has talked about notes, ignoring what was inside of them. While it is true that since its beginning, half century ago, computer music started considering the notes' inside, the prevalent use of concepts like *score* and *orchestra*, common to most Music *N* languages, did certainly not encourage the merger of these two temporal levels (i.e. a *macro* level dealing with notes and a *micro* level dealing with the sound within these notes). But nowadays, the fact is that the separation of both worlds, when applied to computer music creation, is not only becoming less and less trivial, especially since real-time synthesis has become widely available, but can also be considered as a burden and an anachronism.

With the advent of MIDI, almost two decades ago, computer assisted music creation (as opposed to computer assisted sound creation) was able to become a real-time activity, perverting musical concepts such as composition, interpretation and improvisation. But it was not until these very last years that the increasing power of personal computers has allowed the definitive bloom of real-time software synthesizers. Even though virtual synthesizers do not really represent any new concept, they have important differences with their hardware counterparts, as they have the potential to achieve what no manufacturer could ever think of, giving more room for freedom, experimentation and imagination, and bringing forth new sonic and control paradigms.

This is the vast and vague arena in which musical interface inventors and performers have to evolve nowadays. While the more composition-oriented musicians will tend to the macro-level idea, others will favour the more performance-oriented micro-level one, and some will try to find an uncertain equilibrium between both conceptions. My background as an amateur free-jazz saxophonist in the early 1980's, may surely be one of the reasons for which I prefer this no man's land third territory.

**Controllers and Generators**

The separation between gesture controllers and sound generators standardized by MIDI, boosted the creation and development of many new interesting alternative controllers. However, this separation should not always be seen as a virtue or an unavoidable technical imposition. Before MIDI, the term *musical instrument* always referred to both sides of the sound and music creation process (i.e. controller <u>and</u> generator), and it is obvious that the final expressiveness and richness of any musical interface (controller) is not independent of the generator involved and the mapping applied between them. Some of the consequences that result from this fact are truisms (e.g. not all combinations of controllers-generators are equally satisfying), but there are also some important questions that raise: *Are today's standard music communication protocols (e.g. MIDI) wide and flexible enough, or are the potential dialogues between controllers and generators limited or narrowed by these existing protocols? Is it possible to develop highly sophisticated controllers without a prior* *knowledge of how the sound or music generators will work*?

Bi-directional mappings may be one of the solutions leading to wider and richer communications, as they allow haptic or force feedback, and they can exist within the MIDI technology. But force feedback is not the only imaginable feedback: traditional instruments, for example, resonate (i.e. they are "conscious" of the sound they produce) and this acoustic feedback is in fact fundamental in the overall sound production process. While audio analysis, which could be extended to the concepts of music analysis or *machine-listening*, is a regular topic in many computer-music disciplines, this task is not usually undertaken by the controller subsystem.

On the other hand, low-cost and widely available input devices such as mice or joysticks do not necessarily have to be considered as an all-gone dead-end, as there is still a lot of research to be done in the area of interactive GUI. We will see how in the case of the mouse-driven FMOL instrument [4 and 5], the controller audio feedback (presented in its visual form) intuitively helps the understanding and the mastery of the interface, enabling the simultaneous control of a high number of parameters in a manner that could not be possible without this feedback.

### Individual vs. Collective Performance - Dilettante vs. Professional Performers

Creating music with new interfaces poses also new questions about whom and how will be using them. While performing music has typically been a collective event, traditional musical instruments have been mostly designed for an individual use (even if some, as the piano or the drumkit can be easily used collectively). This restriction can now be freely avoided when designing new interfaces, which leads to a new generation of distributed instruments, with a plethora of possible different models following this paradigm (statistical control, equally-allowed or role-playing performers, etc.). Implementations of musical computer networks date back to the late 1970s with performances by groups like the League of Automatic Music Composers or the Hub [1]. This may also be the common case of many interactive sound installations, which respond to the public movements and gestures, which leads us to another important point: that of the skills and the know-how of the performer(s).

I have developed several computer-based interactive music systems since 1989. Some of them were conceived for trained musicians or even for specific performers, while others were to be controlled by members of an audience in public performances. The demands for the two genres are obviously different. Complicated tools, which offer great freedom, can be built for the first group, while the second group demands simple but appealing tools that -while giving their users the feeling of control and interaction- produce "satisfactory" outputs. These two classes are often mutually exclusive. Musicians become easily bored with the "popular" tool, while the casual user may get lost with the sophisticated one. But is this trend compulsory? *Isn't it possible to design interfaces that can appeal to both sectors -tools that would not dishearten hobbyist musicians, but that would still be able to produce completely different musics, allowing a rich and intricate control and offering various stages of training and different learning curves?*

There is the common belief that more hard-to-play instruments lead to richer and more sophisticated musics (e.g. the piano vs. the kazoo)[1], but expressiveness does not really imply difficulty, and in that sense, one of the obvious research trends in musical interfaces design can be the creation of easy-to-use and, at the same time, sophisticated and expressive instruments. The best way to understand and appreciate any discipline, whether artistic or not, and music is no exception, is by doing and being part of it. More *efficient* instruments that can subvert the previous effort-result statement will bring new sophisticated music creation possibilities to non-trained musicians. Lets try, as Robert Rowe suggests, to "develop *computer musicians* that do not just play back music *for* people, but become increasingly adept at making new and engaging music *with* people, at all levels of technical proficiency" [8].

### PREVIOUS WORKS

Influenced by the works and writings of George Lewis [7] and Robert Rowe, *PITEL* (1989-1991) is the first real-time interactive music system I developed [3]. It is a software environment for polyphonic real-time composition and improvisation, which employed some simple mathematical machine-listening techniques. It could generate, under the control of a *mouse-conductor,* up to eight MIDI voices, while listening and reacting to one or two external MIDI players. PITEL performances were usually jam sessions that involved three musicians: a saxophone and a trumpet player, both fitted with pitch-to-MIDI converters, and myself *conducting* the program. All high-level parameters were adjusted by simple means of virtual buttons and sliders, and the instrument remained at the MIDI note level, without attempting any sonic control. It also lacked any serious interface design and conception (which is not necessarily the case of all mouse-driven GUI).

PITEL algorithmic approach makes it very inertial, and fast changes are hard to obtain. In contrast, the *QWERTYCaster*

---

[1] This question although somehow naïve, has nothing to do with the obvious existence of many "simple" but interesting and "rich" musics.

(1996) is a very simple and low-tech guitar-shaped electronic instrument which I designed solely for myself and used for free improvisation, during 1996 and 1997. The aim was to have a fast system with direct audio output in response to every movement or gesture, not unlike Nicolas Collins "Trombone" [2] or Michael Waisvisz "Hands" [6]. As shown in figure 1, it consisted of a QWERTY computer keyboard ("strings"), a trackball ("frets") and a joystick pad ("lever"), all held together in a guitar-like piece of wood. Its five continuous controllers (two degrees of freedom for the trackball and three for the joystick), together with the joystick trigger, buttons and keys, steered an old 486 computer with an sampler soundcard and a simple but effective custom MIDI software. As opposed to PITEL, the QWERTYCaster only focused in sound control, leaving all the macro-formal control and organization for the human player.

*AFASIA* (1998), my third and last collaboration with the visual artist and performer, Marcel.lí Antúnez, is a one-man-show multimedia play inspired in Homer's Odyssey, in which the performer-conductor (Antúnez), fitted with a plethora of radio-emitting sensors, controls with his movements, a whole musical robot orchestra (consisting on a seventy-fingers electric guitar+bass, a walking drumkit, a one-string electric violin and a three-bagpipe "horn section") and the interactive multimedia and DVD video projections (plus an audio CD, a sampler, two audio effects racks, a Yamaha MIDI-controlled audio mixer, a DMX light table, and the video projector input-switcher). The complexity of this setup as opposed to the limited semantics of the sensors employed (gloves, buttons, potentiometers in each of the performer's articulation and mercury switches in his extremities) makes Afasia a peculiar example of score-driven interactive setup, in which every "island" (taken from Ulysses' journey) behaves as an independent interactive environment with different mappings and different restricted degrees of freedom. Afasia and Epizoo have been performed several hundreds of times in more than twenty-five countries and have received international awards. Figures 3 and 4 show Antúnez, in two moments of an Afasia performance.

### FMOL (F@ust Music On-Line)

Far less spectacular than previous work, I still consider FMOL (1997-2000) my deepest experimentation into the new musical expression possibilities. This project started when the Catalan theatre group La Fura dels Baus, proposed me to develop an Internet-based music composition system that could allow cybercomposers to participate in the creation of the music for their next show, F@ust 3.0. Initially I did not have a clear idea of what I wanted; I just knew what I did not want: to allow people to compose tempered music on the keyboard and send us attached MIDI files via E-mail. Besides, although I felt that the project should have a fairly "popular" approach, and did not want to be too demanding and restrictive about the participants' gear, I was not looking for a dull General MIDI sound, but for richer sounds and textures, that could introduce newcomers into more experimental electronic music [2]. Real-time mouse-driven software synthesis seemed therefore the natural solution.

### FMOL's Interface Design

The conception of Bamboo, FMOL's main graphical mouse-controlled interface, was very tight with the synthesis engine architecture design. Both were in fact developed in parallel with the primary aims of conceiving a real-time composition and synthesis system, appealing to both trained electronic composers and more casual or hobbyist musicians, suitable for the Internet (browser plugin, small scorefiles, etc.), that could run on a standard computer fitted with any conventional multimedia soundcard [3]. Discussing FMOL's architecture would be too long, but I will mention that the sound engine supports six stereo real-time synthesized audio tracks, each one consisting of a sound generator (sine, square, sample player, etc.) and three serial processors (filters, reverbs, resonators, etc.), which can be chosen by each composer between more than 100 different synthesis algorithms or variations. Each generator or processor contains in its turn, four independent low frequency oscillators (LFOs), with controllable frequency, range and shape (sinusoidal, square, triangular, saw tooth or random), which can modulate any of the generator/processor parameters. The important point is that all this complicated architecture is clearly reflected in a intuitive, symbolic and non-technical way, in Bamboo's graphical interface.

The Bamboo screen, as shown in figure 4, presents a lattice in which vertical lines are associated with the synthesis generators and horizontal lines with the synthesis processors. Like a virtual guitar, these vertical lines/strings can be plucked or fretted with the mouse while they continuously draw the sound they generate like a multichannel oscilloscope. Horizontal segments, on the other hand, control the synthesizer's serial processors, and can be dragged and oscillate up and down. A Bamboo

---

[2] Other Internet related sites with collaborative music:

- MIT's Brain Opera mixes online audience participatory compositions with live performance:
  http://lethe.media.mit.edu/first-page.html
- William Duckworth's Internet based Cathedral piece:
  http://www.monroestreet.com/Cathedral/main.html
- ResRocketSurfer is a compositional environment and midi collaboration software: http://www.resrocket.com/

[3] Only a Windows version has been produced.

user's manual can not be included but I will just mention that the combination of both mouse buttons and the computer keyboard allows for an intricate control, including sustaining sounds, modifying any parameter, recording gestures loops, applying LFOs with frequency, amplitude and wave-shape control, creating arpeggios or recording and restoring sequences and screen snapshots. Although it may sound quite complex, the abstract visual feedback of all the instrument activity has proved to be an invaluable help for users.

**Musical and Social Implications**

Unlike any other system I have designed, FMOL has been used by hundreds of Internet composers. From January to April 1998, the FMOL first Internet-database received more than 1,100 brief pieces by around 100 composers, some of whom connected nightly and spent several hours a week creating music. One of our main goals (i.e. to conceive a musical system which could be attractive to both trained and untrained electronic musicians) was fully attained. We know now that several of the participants had no prior contact with experimental electronic music and a few were even composing for the first time, but all of them took it, however, as a rather serious game, and the final quality level of the contributions was impressive. After a difficult selection process (only 50 short pieces could be chosen and included on the show's soundtrack), and considering that a great number of interesting compositions had to be left aside, we decided some months later to produce a collective CD with a mixture of old and new compositions.

A new web with a new version of the software has been back on-line during September 2000 for La Fura´s new show, the opera DQ, premiered last October at the Gran Teatre del Liceu in Barcelona[4]. During one month, more than 600 compositions have been submitted, and the selected ones constitute now the electroacoustic parts of an otherwise orchestral score.

In September 2000, a one-week workshop specially aimed at visual artists took place in Lisbon. The workshop concluded with several conducted collective improvisations, with astonishing results.

Since 1999 FMOL has also become my main instrument at live concerts. The FMOL Trio (Pelayo Arrizabalaga / saxophones and clarinets, Cristina Casanova / FMOL and myself) performs improvised electronic music, while two projectors connected to each of the computers give the complementary visual feedback, enabling the audience to <u>watch the music and how it is being constructed</u>. Figure 5 shows the group in concert. A live CD has been released this year[5].

**FUTURE WORK**

Further developments of the FMOL instrument have taken different courses (as a model for Internet collective music creation and as a professional controller specially oriented to performance). Along these last line, we are developing a "King Size concert version", which will use video detection in order to allow the performers to interact directly with their hands, over a 3x2 meters retro-projected bamboo screen.

Besides these ongoing projects, at the Music technology Group of the Pompeu Fabra University (already very active in the area of sound analysis and synthesis), we are initiating an Interactive Systems research line with several projects starting this year.

**REFERENCES**

1. Bischoff, J., Gold, R., and Horton, J. Music for an Interactive Network of Computers. *Computer Music Journal, Vol. 2, No.3, 1978*.
2. Collins, N. Low Brass: Trombone-Propelled Electronics. *Leonardo Music Journal, Vol. 1, No. 1, 1991*.
3. Jordà, S. A Real-Time MIDI Composer and Interactive Improviser by Means of Feedback Systems. *ICMC Proceedings, 1991*.
4. Jordà, S. A graphical and net oriented approach to interactive sonic composition and real-time synthesis for low cost computer systems. *Digital Audio Effects Workshop Proceedings, 1998*.
5. Jordà, S. Faust music On Line: An Approach to Real-Time Collective Composition on the Internet. *Leonardo Music Journal, Vol. 9, 5-12, 1999*.
6. Krefeld, V. The Hand in The Web: An Interview with Michel Waisvisz, *Computer Music Journal, Vol. 14, No. 2, 1990*.
7. Roads, C. Improvisation with George Lewis. In *Composers and the Computer*, ed. C. Roads. Los Altos, Calif., William Kaufmann, Inc, 1985.
8. Rowe, R. *Interactive Music Systems – Machine Listening and Composing*, The MIT Press, Mas, 1992. p. 263.

---

[4] Visit the web, download the software or learn more about the DQ-FMOL project at http://teatredigital.fib.upc.es/dq

[5] MP3 audio and video excerpts from the Internet projects, the FMOL Trio or the Lisbon workshop can be accessed at http://www.iua.upf.es/~sergi/download/chi2001. More information about all these projects is also available at the author's homepage http://www.iua.upf.es/~sergi and at http://www.arrakis.es/~ccs/ktonycia

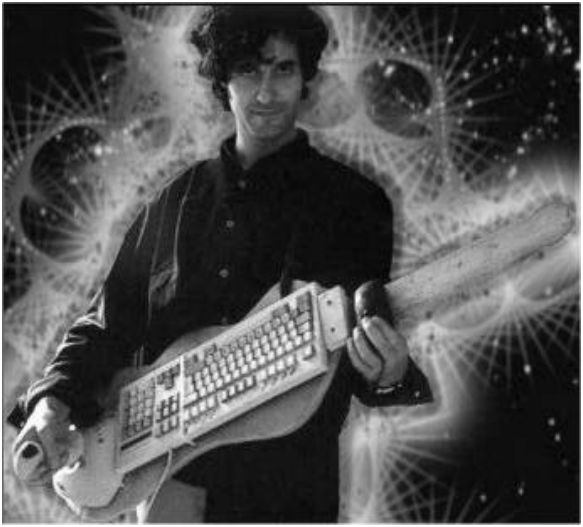

Figure 1. Sergi Jordà with the *low-tech QWERTYCaster* (1996)

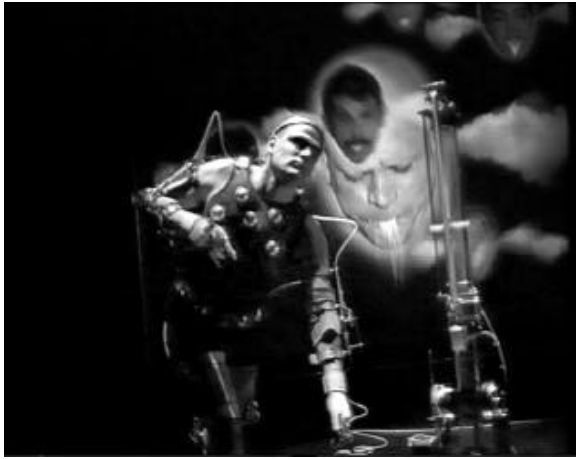

Figures 2 and 3. Marcel.lí Antúnez controlling the violin robot, and the guitar robots in Afasia (1998)

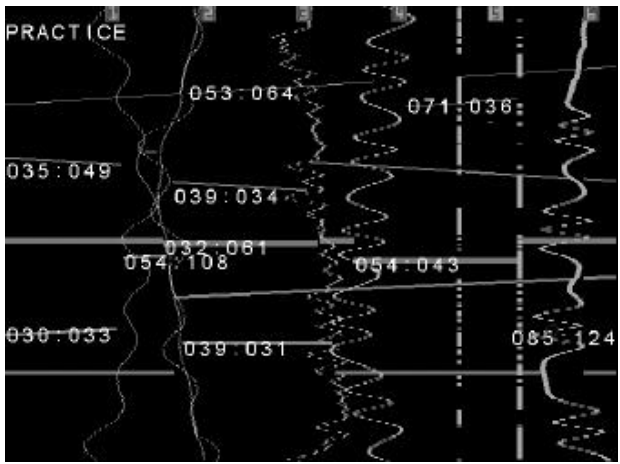

Figure 4. Screenshot of Bamboo, FMOL graphical interface, in full action (1999).

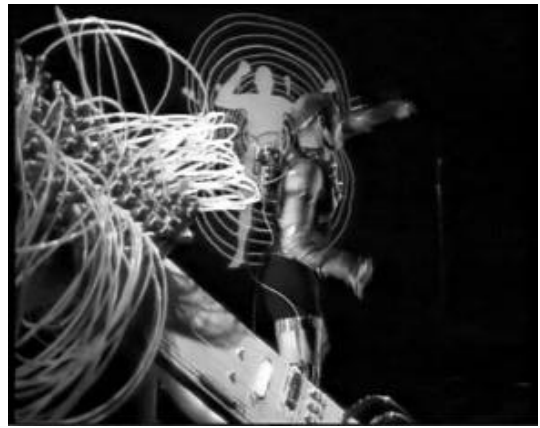

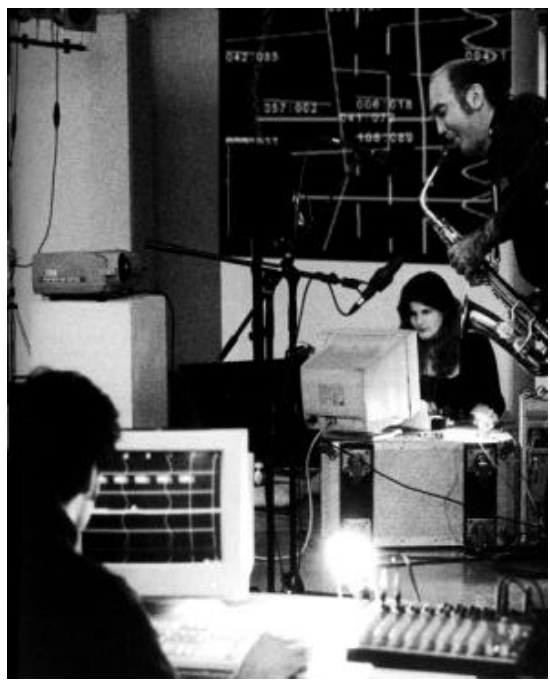

Figure 5. The FMOL Trio in performance. From left to right: Sergi Jordà (FMOL computer), Cristina Casanova (FMOL computer), Pelayo F. Arrizabalaga (alto sax) (2000).